\documentclass[pdflatex,sn-aps,iicol]{sn-jnl}

\usepackage{graphicx}%
\usepackage{multirow}%
\usepackage{amsmath,amssymb,amsfonts}%
\usepackage{amsthm}%
\usepackage{mathrsfs}%
\usepackage[title]{appendix}%
\usepackage{xcolor}%
\usepackage{textcomp}%
\usepackage{manyfoot}%
\usepackage{booktabs}%
\usepackage{algorithm}%
\usepackage{algorithmicx}%
\usepackage{algpseudocode}%
\usepackage{listings}%
\usepackage{dcolumn}
\usepackage{bm}
\usepackage{hyperref}
\usepackage{multirow}
\usepackage{gensymb}

\begin{document}

\title[]{X-ray pulsed light curves of highly compact neutron stars  as probes of scalar-tensor theories of gravity}

\author*[1,2]{\fnm{Tulio} \sur{Ottoni}}\email{tulio.costa@edu.ufes.br}

\author[3]{\fnm{Jaziel} \sur{G. Coelho}}\email{jaziel.coelho@ufes.br
}
\equalcont{These authors contributed equally to this work.}

\author[4]{\fnm{Rafael} \sur{C. R. de Lima}}\email{rafael.nuclear@gmail.com}
\equalcont{These authors contributed equally to this work.}

\author[3,5,6]{\fnm{Jonas} \sur{P. Pereira}}\email{jpereira@camk.edu.pl}
\equalcont{These authors contributed equally to this work.}

\author[2,7,8,9,10]{\fnm{Jorge} \sur{A. Rueda}}\email{jorge.rueda@icra.it}
\equalcont{These authors contributed equally to this work.}

\affil*[1]{\orgdiv{PPGCosmo}, \orgname{Universidade Federal do Espírito Santo}, \orgaddress{\street{Av. Fernando Ferrari}, \city{Vitória}, \postcode{29075-910}, \state{Espirito Santo}, \country{Brazil}}}

\affil[2]{\orgdiv{ICRANet-Ferrara, Dipartimento di Fisica e Scienze della Terra}, \orgname{Universit\`a degli Studi di Ferrara}, \orgaddress{\street{Via Saragat 1}, \city{Ferrara}, \postcode{I--44122}, \state{FE}, \country{Italy}}}

\affil[3]{\orgdiv{Núcleo de Astrofísica e Cosmologia (Cosmo-Ufes) \& Departamento
de Física}, \orgname{Universidade Federal do Espírito Santo}, \orgaddress{\street{Av. Fernando Ferrari}, \city{Vitória}, \postcode{29075-910}, \state{ES}, \country{Brazil}}}

\affil[4]{\orgdiv{Departamento de Física}, \orgname{Universidade do Estado de Santa Catarina}, \orgaddress{\street{Rua Paulo Malschitzki}, \city{Joinville}, \postcode{89219-710}, \state{SC}, \country{Brazil}}}

\affil[5]{\orgdiv{Departamento de Astronomia, Instituto de Astronomia, Geofísica e Ciências Atmosféricas (IAG)}, \orgname{Universidade de
São Paulo}, \orgaddress{\street{Rua do Matão 1226, Cidade Universitária}, \city{São Paulo},  \postcode{05508-090}, \country{Brazil}}}

\affil[6]{\orgdiv{Nicolaus Copernicus Astronomical Center}, \orgname{Polish Academy of Sciences}, \orgaddress{\street{Bartycka 18}, \city{Warsaw}, \postcode{00-716}, \country{Poland}}}

\affil[7]{\orgdiv{ICRANet}, \orgaddress{\street{Piazza della Repubblica 10}, \postcode{I-65122}, \state{Pescara}, \country{Italy}}}

\affil[8]{\orgdiv{ICRA, Dipartamento di Fisica}, \orgname{Sapienza Università di Roma}, \orgaddress{\street{Piazzale Aldo Moro 5},\postcode{I-00185}, \city{Rome}, \country{Italy}}}

\affil[9]{\orgdiv{Department of Physics and Earth Science}, \orgname{University of Ferrara}, \orgaddress{\street{Via Saragat 1},\postcode{I-44122}, \city{Ferrara}, \country{Italy}}}

\affil[10]{\orgdiv{INAF}, \orgname{ Istituto di Astrofisica e Planetologia Spaziali}, \orgaddress{\street{Via Fosso del Cavaliere 100},\postcode{ I-00133}, \city{Rome}, \country{Italy}}}

\abstract{The strong gravitational potential of neutron stars (NSs) makes them ideal astrophysical objects for testing extreme gravity phenomena. We explore the potential of NS X-ray pulsed light-curve observations to probe deviations from general relativity (GR) within the scalar-tensor theory (STT) of gravity framework. We compute the flux from a single, circular, finite-size hot spot, accounting for light bending, Shapiro time delay, and Doppler effect. We focus on the high-compactness regime, i.e., close to the critical GR value $GM/(c^2 R)=0.284$, over which multiple images of the spot appear and impact crucially the lightcurve. Our investigation is motivated by the increased sensitivity of the pulse to the scalar charge of the spacetime in such high compactness regimes, making these systems exceptionally suitable for scrutinizing deviations from GR, notably phenomena such as spontaneous scalarization, as predicted by STT. We find significant differences in NS observables, e.g., the flux of a single spot can differ up to 80\% with respect to GR. Additionally, reasonable choices for the STT parameters that satisfy astrophysical constraints lead to changes in the NS radius relative to GR of up to approximately 10\%. Consequently, scalar parameters might be better constrained when uncertainties in NS radii decrease, where this could occur with the advent of next-generation gravitational wave detectors, such as the Einstein Telescope and LISA, as well as future electromagnetic missions like eXTP and ATHENA. Thus, our findings suggest that accurate X-ray data of the NS surface emission, jointly with refined theoretical models, could constrain STTs.}

\keywords{Pulsar Pulse Profile, Scalar-Tensor Theory, X-ray Lightcurve}



\maketitle

\section{Introduction}\label{sec1}

Neutron stars (NSs) are natural laboratories for testing fundamental physics, ranging from interactions above nuclear saturation density to the strong gravitational field in the stellar interior and surroundings. Their astrophysical observations can probe fundamental interactions in a very unique regime \cite{Nattilla2022,Berti2015}. Regarding the gravitational field, the extreme conditions of density and pressure in NSs can activate non-minimally coupled fields to gravity. The simplest case is that of scalar fields. In the context of scalar-tensor theories (STTs), this is one way to understand the phenomenon of \textit{spontaneous scalarization}, a novel non-perturbative effect arising in these theories. This effect predicts deviations from General Relativity (GR) that can be observationally tested \cite{FujiiMaeda2003, Doneva2022,damour92,damour93,damour96PhysRevD.54.1474}. Scalar fields are pivotal in cosmological scenarios, leading to well-established inflationary models \cite{Faraoni:2000gx,Clifton:2011jh,AndersonYunesPhysRevD.94.104064}. A fundamental scalar field can modify compact objects' structure and gravitational field depending on the Lagrangian coupling between the scalar field and ordinary matter. Such theoretically predicted modifications can be tested with astrophysical observations. On the other hand, some grand unification theories, like string theory, also predict scalar degrees of freedom in the low energy, classical regime (see, e.g., \cite{Hollands:2014eia,Polchinski:1998rq}, and references therein). 

We aim to extend previous efforts to constrain STTs using astrophysical observations of NSs, particularly the X-ray light-curve profiles. This task can generally be done for STTs without specifying the origin of the scalar field. There is currently a broad class of astrophysical sources associated with NSs, with observations spanning the electromagnetic spectrum from the radio to high-energy X-rays and gamma rays. In particular, some systems exhibit periodic X-ray emissions modulated by the star’s rotation period, which can be deduced from the pulsar timing of radio signals. The modeling of these pulse profiles can be compared to observations, such as those made by NICER \cite{Nicerdescriptionpaper}, and global properties of NSs, like compactness (${\cal C}\equiv GM/Rc^2$, where $M$ is the NS mass and $R$ its radius), can be inferred from the analysis. This, in turn, can constrain the equation of state of nuclear matter. Furthermore, modeling deviations from GR in the pulse profile can, in principle, constrain modified gravity models, such as STT. 

A crucial and pertinent question regarding NSs revolves around the maximum mass allowed by gravitational instability, known as the Tolman-Oppenheimer-Volkov (TOV) limit. Presently, we have reliable mass measurements for high-mass pulsars, with masses around or greater than $2M_\odot$, such as PSR J0348+0432 with $2.01 \pm 0.04M_\odot$ \cite{Antoniadisscience.1233232}, PSR J0740+6620 with $2.08 \pm 0.07M_\odot$ \cite{Fonseca_2021} and PSR J0952-0607 with $2.35 \pm 0.17M_\odot$ \cite{2022ApJ...934L..17R}. The first two pulsars are in a binary system, and the masses were estimated by standard timing techniques. At the same time, the last one is a ``spider" system with more model-dependent uncertainty. But the message here is that such massive NSs are feasible, and such systems are ideal for testing strong gravity effects because of the high gravitational binding energy compared to low-mass NSs. 

For a $\sim2M_\odot$ star, several realistic equations of state (EOS) predict a radius such that the resultant compactness can be closer or higher than $GM/Rc^2=0.284$, the critical value in GR that makes light bending strong enough for the whole NS surface to be seen by an observer at rest at infinity \cite{Sotani3PhysRevD.98.044017}. As a result, considering multiple images is relevant to model realistic lightcurves of high-mass stars. From the point of view of STT predictions, deviations from GR generally increase with compactness, making these high-mass systems ideal for testing and constraining the strong field regime of alternative theories. This study focuses on scenarios of high compactness, which have been partially explored in the literature and are critical for understanding the differences between STT and GR. Our analysis emphasizes the effects of possible compactness values approaching the theoretical limits.

Significant research has been conducted on pulse profile modeling within the framework of STT \cite{Sotani1PhysRevD.96.104010,Sotani2PhysRevD.96.104018, SilvaeNunesPhysRevD.99.044034,XuGaoShaoPhysRevD.102.064057,HuGaoXuShaoPhysRevD.104.104014}. Silva and Yunes \cite{SilvaeNunesPhysRevD.99.044034} derived the flux of infinitesimal spots, incorporating the varying effects of bending, time delay, and kinematic factors specific to STT. Here, we use their expression but integrate it over a finite spot and a different regime of compactness. Furthermore, the work by \cite{XuGaoShaoPhysRevD.102.064057,HuGaoXuShaoPhysRevD.104.104014} expanded these calculations to finite spots and linked them to particular scalar-tensor models with a massive scalar field. In this study, we investigate the impact of extended spots on the lightcurve of an isolated NS with high compactness \cite{Sotani3PhysRevD.98.044017,Sotani4PhysRevD.101.063013}, demonstrating that such compact systems, close to producing multiple images of the spot by a strong lensing effect, are promising for testing deviations from GR, as evidenced by the qualitative and quantitative differences in the light curves.

The structure of the paper is as follows. In Section \ref{stt}, we review the fundamentals of Scalar-Tensor theory and examine specific models that predict spontaneous scalarization. Section \ref{pulse profile} reviews pulse profile modeling techniques within the STT context. Finally, in Section \ref{results}, we present and discuss our findings. Throughout this paper, we adopt the units where $G=c=1$.

\section{Scalar-Tensor theory} \label{stt}

A general class of scalar-tensor theory that encodes a nonminimal coupling with geometry is described by the gravitational action
\begin{multline}
    S_g=\frac{1}{16 \pi}\int d^4x \sqrt{-\tilde{g}}[F(\Phi)\tilde{R} - \\ - Z(\Phi) \nabla_\mu\Phi\nabla^\mu\Phi - V(\Phi)],
\end{multline}
which is written in the so-called \textit{Jordan} Frame, where the scalar field couples directly with the geometry via $F(\Phi)$ \cite{Sotiriou:2007zu}. We get a particular theory within this general class once we specify a particular form of these functions.

In this work, we focus on the simpler case of a massless scalar field with no self-interactions so that we can neglect the potential term $V(\Phi)=0$. Also, to take into account the stringent constraints from solar system experiments \citep[see, e.g.,][]{cassinibound}, we set the background scalar field value to $\Phi_\infty=0$ since parametrized pós Newtonian (PPN) deviations in this theory are generally proportional to this background value. Including the matter contribution, the total action reads
\begin{equation}
    S = S_g + S_m[\Psi_m,\tilde{g}_{\mu\nu}],
\end{equation}
where $\Psi_m$ denotes the matter fields collectively. As usual, we assume a perfect fluid form for the energy-momentum tensor $T^{\mu\nu} \equiv (2/\sqrt{-g})\delta S_m/\delta \tilde{g}_{\mu\nu}$, i.e.,
\begin{equation}
    T^{\mu\nu}=(\epsilon + p)u^\mu u^\nu + p\tilde{g}^{\mu\nu}.
\end{equation}
For numerical computation, it is convenient to work in the so-called \textit{Einstein} frame, where, after a conformal transformation of the metric $g_{\mu\nu}\equiv F(\Phi)\tilde{g}_{\mu\nu}$ and field redefinition $\Phi \rightarrow \varphi(\Phi)$, the action can be written as
\begin{multline}
    S=\frac{1}{16 \pi}\int d^4x \sqrt{-g}[R - 2 \nabla_\mu\varphi\nabla^\mu\varphi] + \\  S_m[\Psi_m,A^2(\phi)g_{\mu\nu}],
\end{multline}
where we define $A(\varphi)\equiv F(\Phi(\varphi))^{-1/2}$ and now the scalar field is minimally coupled to gravity, but it couples directly with matter. We also chose the standard canonical kinetic term $Z(\Phi)=1$. After varying the action with respect to $g_{\mu\nu}$ and $\varphi$, we get the following field equations:
\begin{gather}
    G_{\mu\nu} - 2\partial_\mu\varphi_\nu\partial\varphi + g_{\mu\nu} g^{\alpha\beta}\partial_\alpha\varphi\partial_\beta\varphi = 8\pi T_{\mu \nu} A^2(\varphi). \\
\nabla^\mu\nabla_\mu\varphi = -4\pi \label{scalarfieldeq}A^4(\varphi)\alpha(\varphi)T,
\end{gather}
where $T\equiv g_{\mu\nu}T^{\mu\nu}=3p-\epsilon$ is the trace of the energy-momentum tensor,
\begin{equation}
    \alpha(\varphi)\equiv \frac{d \ln A(\varphi) }{d\varphi}.
\end{equation}

\subsection{Spontaneous Scalarization}

The most distinct feature of the STT when compared to GR is the phenomenon of \textit{spontaneous scalarization} of compact objects, discovered by Damour and Esposito-Farèse \cite{damour93}. This scalarization is a violation of the strong equivalence principle associated with a gravitational phase transition \cite{Doneva1PhysRevLett.129.121104,Doneva2PhysRevD.108.044054}. It can be understood as a tachyonic instability of the scalar field \cite{tachyonicPhysRevD.96.064009}. To see this more clearly, we can study linear perturbations of the scalar field  given by Eq. (\ref{scalarfieldeq}),
\begin{equation}
    \square\delta\varphi=-4\pi \beta(\varphi) T\delta\varphi,
\end{equation}
where $\beta(\varphi)=d\alpha(\varphi)/d\varphi|_{\varphi_\infty}$. This equation is a Klein-Gordon equation of the background space-time, with an effective mass
\begin{equation}
    \mu_{eff}^2 = -4\pi \beta T. 
\end{equation}
The solutions are oscillatory for a positive effective mass squared, and the perturbations do not grow. This happens if $\beta$ and $T$ have opposite signs. But now, if they have the same sign, the instability grows until the linear approximation breaks down and the nonlinearities occur, quenching the scalar field's growth.

\subsection{Models} \label{models}

Once we choose a specific form for the coupling function, we select a particular model within the general class of STTs. A simple model is described by an exponential coupling, first used in \cite{damour93}
\begin{equation}
    A(\varphi)=e^{\frac{\beta\varphi^2}{2}},
\end{equation}
frequently known as Damour-Esposito-Faresè (DEF) theory and has a significant historical value and simplicity, although incompatible with recent observations \cite{Zhao2022}.
Another well-motivated form for the conformal factor comes from cosmology, especially from inflationary models \cite{Faraoni1PhysRevD.53.6813,InflationPhysRevD.40.1753}
\begin{equation}
    A(\varphi) = \frac{1}{\sqrt{1 + \xi \Phi^2}},
\end{equation}
but the technical difficulty here is that we need to solve the relation between the fields numerically, and so there is no close form for $\alpha(\varphi)$ in the Einstein frame, for example, \cite{nonminimalPhysRevD.99.064055}. This difficulty can be overcome with the use of an analytical approximation using hyperbolic functions, where the conformal function is
\begin{equation}
    A(\varphi) = (\text{cosh}(-2\sqrt{3}\xi\varphi))^{-      \frac{1}{6\xi}},
\end{equation}
while the coupling function is
\begin{equation}
    \frac{1}{\sqrt{3}}\tanh (-2\sqrt{3}\xi \varphi).
\end{equation}
This model was first discussed in \cite{mandesandortizPhysRevD.93.124035}, known as the Mendes-Ortiz (MO) theory. Finally, all three models are similar for $\xi=2\beta$, showing the same linear behavior when expanded in powers of $\phi$.

\subsection{Exact External Solution}

In the scalar-tensor theory, an exact analytical solution for a spherically symmetric spacetime is the Just metric \cite{Just,Coquereaux1990qs,Creci2023}. Written in the Einstein frame, it is
\begin{equation}\label{metric}
    ds^2 = -f^{b/a}dt^2 + f^{-b/a}d\rho^2 + \rho^2f^{1-b/a}d\Omega,
\end{equation} 
besides the spherical part, the radial coordinate $\rho$ is related to Schwarzschild coordinate by
\begin{equation} \label{radialcoordinates}
    r=\rho(1-a/\rho)^{(1-b/a)/2},
\end{equation}
which cannot be analytically inverted. Here, $b$ is related to the gravitational (ADM) mass, $b\equiv2M$,  and
\begin{equation}
    f\equiv 1 - a/\rho,
\end{equation}
where $a$ has length units and is related to the mass and the scalar field configuration. We recover GR when $a=b$.

The scalar field profile outside the star has the form
\begin{equation} \label{external scalar field}
    \varphi = \varphi_\infty + \frac{q}{a}\text{log}\left(1-\frac{a}{\rho}\right),
\end{equation}
where $\varphi_\infty$ is the background value of the scalar field, which is constrained to be very small by solar system experiments. For simplicity, we assume $\varphi_\infty=0$. Far from the source, the scalar field behaves like an electric field of a point charge $\varphi \sim -q/\rho$, and thus, we make the identification of $q$ as the scalar charge. The constants $a$, $b$ and $q$ are not independent:
\begin{equation} \label{}
    a^2 - b^2 - (2q)^2 = 0.
\end{equation}
It is more common to define a scalar-to-mass ratio $ Q \equiv  q/M$, so that
\begin{equation}\label{constraintcharge}
  a/b = \sqrt{1+Q^2}
\end{equation}
%

\subsection{Constraints}

Since the formulation of GR, it has passed the scrutiny of experimental tests \cite{will_2018}, and tight constraints were put in alternative theories of gravity. The most relevant constraint in the weak field regime and solar system scale comes from the Cassini bound \cite{cassinibound}. Making the PPN expansion for the weak field/low velocities regime, the constraint on the $\gamma$ parameter is usually proportional to the background value of the scalar field. We put $\Phi_\infty=\varphi_\infty=0$, which automatically satisfies that. But even with a restricted weak field phenomenology, STTs can still have a rich, strong-field landscape, bigger than GR, which is precisely the essence of the phenomenon of spontaneous scalarization.

From the strong field perspective, the constraints usually come from the timing of binary systems \cite{moststringenttest, doublepulsar2021PhRvX..11d1050K,2024FreireandWex}. These bounds are typically put in the microphysics of the theory, i.e., the restrictions on coupling parameters, which, in the case of the models discussed before, translates into a constraint on the coupling constant $\xi$. Indeed, the timing of radio pulsars in binary systems leads to the exclusion of the region $2\xi=\beta\lesssim-4.35$, mainly due to the effect of emission of dipolar gravitational radiation, which affects the dynamics of the two-body system at 1.5 post-Newtonian order (PN), in contrast with the quadrupolar emission of GR, that enter in a $2.5$ PN order \cite{moststringenttest}.

From the macroscopic phenomenological perspective, few constraints exist on the scalar charge $Q$. Horbatsch and Burgess \cite{chargebound}, with model-independent analysis using the double pulsar, found $Q<0.21$. But it is important to emphasize that a constraint on the scalar charge for a $\sim 1.4 M_\odot$ NS does not necessarily translate into a constraint on the charge of a high mass $\sim 2.1 M_\odot$ because there can be some models that allow spontaneous scalarization only for high mass NSs.

It is important to note also that pulsar timing of binary systems cannot constrain a massive STT when the orbital separation is larger than the characteristic length of the scalar field (Compton wavelength), $\lambda_\phi = (2\pi \hbar)/m_\varphi$. Thus, the scalar field is local, only affecting the NS structure and its immediate surroundings, leaving the orbital motion of wide binaries as in GR \citep[see][]{massivescalarPhysRevD.93.064005}. Another similar case arises when one considers fast rotation \cite{doneva_rapidly_2014, yazadjiev_slowly_2016}, with the strength of the scalar field increasing at the center and inside the star but decreasing quickly after some star radii. 

For a massive STT, the main effect is the suppression of the scalar field proportional to the Yukawa term $e^{-r/\lambda_\phi}$ \cite{massivescalarPhysRevD.93.064005}. We stress that the constraints in the massive scenario differ significantly from those in the massless case. Observationally, the agreement between the orbital motion of close relativistic binaries and GR places constraints on STT, as STT predicts dipolar gravitational wave emission \cite{moststringenttest}. Given that the typical size of such compact binaries is about $r_{\rm bin}\sim 10^{10}m$, dipolar gravitational radiation is suppressed if $r_{\rm bin} > \lambda_{\phi}$, which implies $m_\varphi > 10^{-16}eV$.
On the other hand, a maximum mass can be estimated by ensuring that scalarization is not suppressed inside the star. This condition requires $\lambda_\phi > R$, or equivalently, $m_\varphi < 10^{-9}eV$. Therefore, the allowed mass range for the scalar field is $10^{-16}eV \lesssim m_\varphi \lesssim 10^{-9}eV$, which also accommodates a much broader range of the parameter $\xi$ consistent with observations \cite{massivescalarPhysRevD.93.064005,massivestructure_2020}.

\section{Pulse profile modeling} \label{pulse profile}

Pulse profile modeling is a powerful and crucial tool for analyzing localized surface emission of pulsars, such as the X-ray observations of NASA's NICER observatory \cite{Nicerdescriptionpaper}. Since the seminal work of Pechenick et al. \cite{pechenick}, several studies have been made to model realistic pulse profile of neutrons stars \cite{PoutanenBeloborodov, Cadeau_2007,futureextp2019SCPMA..6229503W}, and by comparing them with data, its possible to infer the mass and radius of the star \cite{Riley:2019yda,Miller:2019cac,2021ApJ...918L..28M,2021ApJ...918L..27R, Choudhury2024}. In addition, the magnetic field structure (related to the spot configurations \cite{Bilous2019}) and hot-spot temperatures could be obtained. 

The basic idea is to make two transformations of the relevant quantities that describe the radiation emission. The first is from a frame co-rotating with the star to a frame just above the star's surface (a local Lorentz boost). The second transformation is from the star to the observer at infinity. The first transformation considers the special relativistic effects of the moving spot, such as aberration and Doppler boost. The second one, being non-local, collects the effect of gravity on photon propagation, such as the bending and time delay.

In the geometric optics approximation, the photon path is a null geodesic of the Just metric (\ref{metric}), whose Lagrangian is
\begin{equation}
    \mathcal{L} = g_{\mu \nu}p^{\mu}p^{\nu},
\end{equation}
where $p^\alpha$ is the photon 4-momenta along the path $x^{\alpha} = (ct,\rho,\theta,\phi)$, parametrized by an affine parameter $\lambda$. The equations of motion follow from Euler-Lagrange equations, with two well-known constants of motion associated with energy and angular momentum conservation
\begin{align}
    dt/ d \lambda & = A^{-2}\epsilon f^{-b/a}, \\
    d\rho/d\lambda & = A^{-4}[c^2\epsilon^2 - (h/\rho)^2 f^{2b/a - 1}], \\
    d\theta / d\lambda &= 0, \\
    d \psi/d\lambda &= A^{-2}(h/\rho^2)f^{b/a - 1}.
\end{align}
We can make the identification of $\psi$ with the azimuthal angle $\phi$ because of the spherical symmetry, i.e., the photon is always constrained to move on a plane, which we can choose as  $\theta=\pi/2$. Also, from the two constants of motion ($\epsilon$ and $h$), we can define, as usual, the impact parameter of the photon as $\sigma\equiv h/\epsilon$.

Now we consider the emission angle $\alpha$ of the photon at the stellar surface, we have tan$\,\alpha = [p^\psi p_\psi/(p^\rho p_\rho)]^{1/2}$ which gives a relation for the impact parameter $\sigma$:
\begin{equation} \label{impact parameter formula}
    \sin \alpha = \frac{\sigma}{\rho_s}(1 - \Bar{a}_s)^{b/a - 1/2},
\end{equation}
where $\Bar{a}_s=a/\rho_s$, and $\rho_s$ is the stellar radius in Just coordinates. Using the geodesic equations and the impact parameter, Silva and Yunes  \cite{SilvaeNunesPhysRevD.99.044034} were able to derive an integral expression for the angle $\psi$ that generalizes the GR expression
\begin{multline}\label{bendingintegral}
    \psi = 2 \sin \alpha \int_0^1 dx \, x[1 - \Bar{a}_s(1-x^2)]^{b/a - 1} \times \\
    \{(1 - \Bar{a}_s)^{2b/a - 1} - (1-x^2)^2 \\ [1 - \Bar{a}_s(1-x^2)]^{2b/a - 1}\sin^2 \alpha\}^{-1/2},
\end{multline}
where $x=\sqrt{1-y}$ and $y\equiv\rho_s/\rho$. (Although we do not use it in this work, in Appendix \ref{belo aprox} we compare the Beloborodov approximation \cite{Beloborodov2002} for GR and its equivalent for STT.) In the GR limit, $a/b=1$, and $\Bar{a}_s$ becomes twice the compactness $2M/R$. In other words, in GR, the bending of the photon path will depend on the emission angle and the compactness, i.e., $\psi_{\rm GR}=\psi_{\rm GR}(\alpha,M/R)$. On the other hand, in STT, the bending will also depend on the scalar charge of the spacetime and the value of $\Bar{a}_s$, i.e., $\psi_{\rm STT} = \psi_{\rm STT}(\alpha, \Bar{a}_s,Q)$

In particular, the visible part of the star is defined by the light ray emitted tangentially to the local radial direction at the star's surface, i.e., $\alpha=\pi/2$. For low compactness, the value of $\psi$ is close to $\alpha$, and half of the star surface is visible. But, as compactness increases, the bending increases, and $\psi$ can become larger than $\pi$, meaning that there is a region behind the star where the light rays emitted can take two different paths to reach the observer, one in the direction of increasing $\psi$ and the other of decreasing $\psi$. (An explicit example of such a phenomenon is given in the Appendix \ref{apB}.) As shown in Fig \ref{fig1}, in GR $\psi_{\rm GR}(\pi/2,M/R)\equiv\psi_c=\pi$ for a compacteness $M/R=0.284$, and the whole star surface is visible. We stress that $\psi_c$ in STT changes nonlinearly with the compactness, as clear from Eq. (\ref{bendingintegral}). Thus, the same should happen with the relative changes of $\psi_c$ and observables in STT and GR that depend on $\psi$.

\begin{figure}
\centering
\includegraphics[width=\columnwidth]{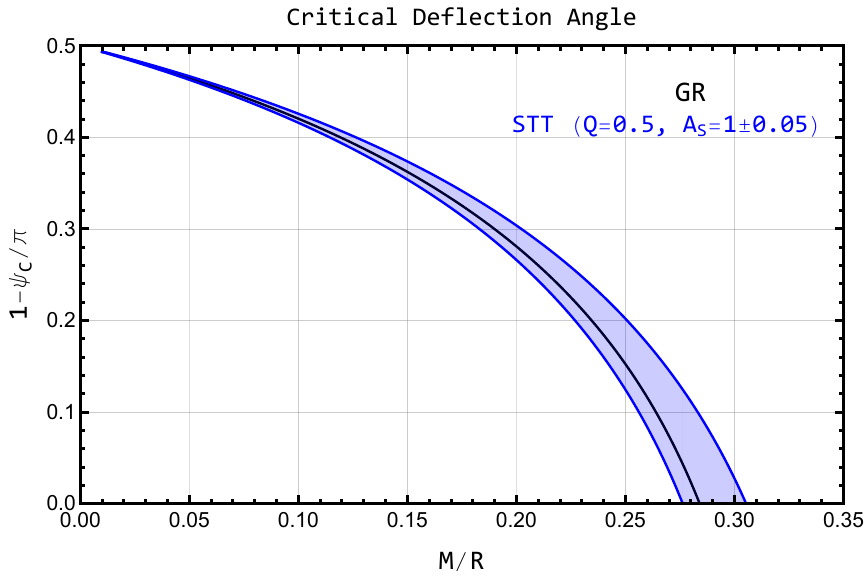}
\caption{Critical deflection angle in GR and Scalar-Tensor theory, with a charge of $Q=0.5$ and a scale factor evaluated on the surface $A_s = 1 \pm 0.05$, where $A_s=1.05$ correspond to the left boundary and $A_s=0.95$ to the right boundary of the blue shaded area. Since the critical angle is equal to $\pi$, which means a star whose surface is fully visible, this affects the lightcurve and could be used to distinguish between the two theories. Notice that the relative changes of the critical view angle in STT and GR scale with the compactness in a nonlinear way.}
\label{fig1}
\end{figure} 

\begin{figure}[h]
\centering
\includegraphics[width=\columnwidth]{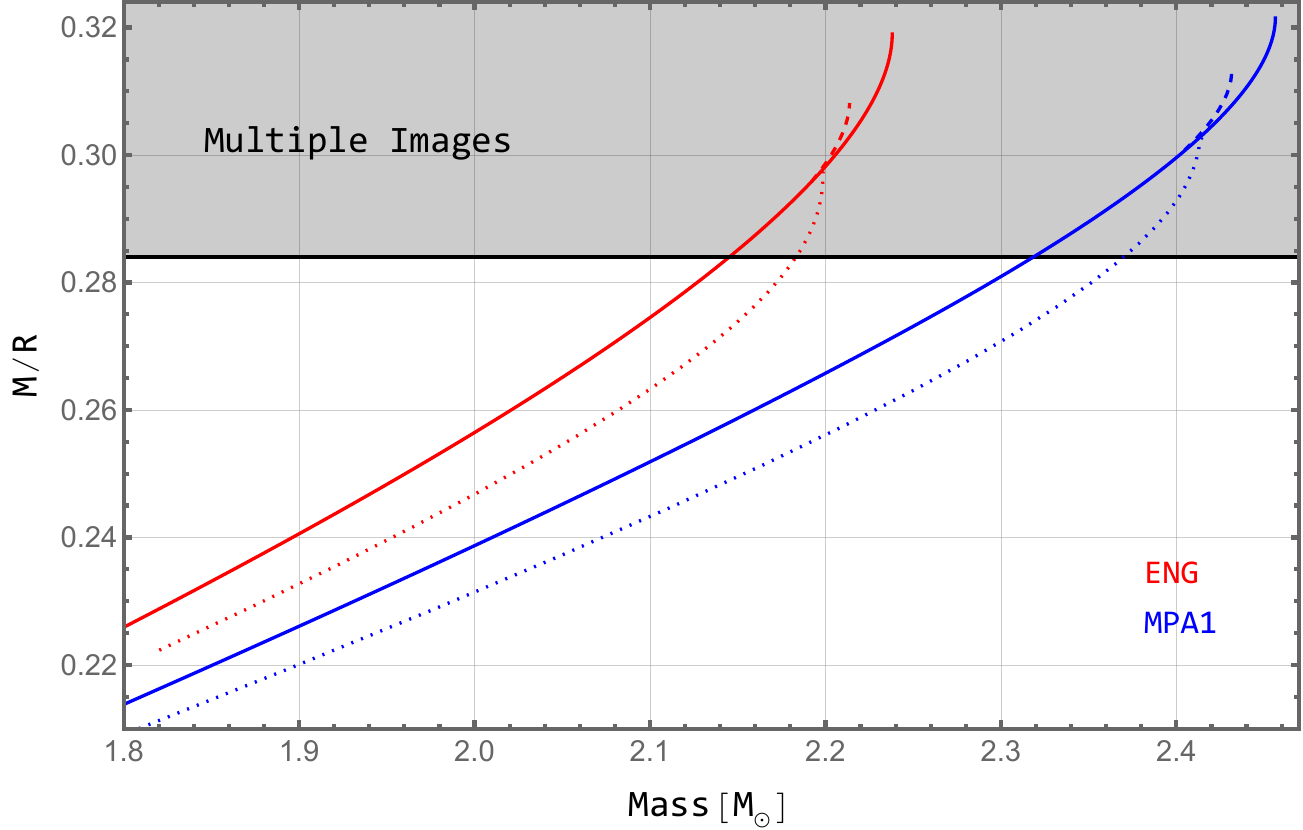}
\caption{Compactness versus ADM mass for two realistic EOS. The continuous lines represent the GR solution. The dotted lines correspond to scalarized solutions in STT for the MO coupling with $\xi=-3$, and the dashed lines at the end of each curve represent the scalarized solution with $\xi=25$. We also mark the critical GR compactness that starts to produce a multiple-image region behind the star ($M/R=0.284$).}
\label{compacteness}
\end{figure}

In addition to the bending, the Shapiro time delay \cite{1983bhwd.book.....S} also affects the photons. Working with the geodesics of the Just metric, Silva and Yunes \cite{SilvaeNunesPhysRevD.99.044034} also derived an integral expression for the time delay, defined as

\begin{equation}
    \Delta t \equiv t(\sigma) - t(\sigma=0),
\end{equation}
i.e., the time difference for a photon emitted directly towards the observer. With all these ingredients, the authors derive the differential flux formula in the context of the STT
\begin{equation} \label{fluxformula}
    dF = A_s^2(1 - \Bar{a}_s)^{\frac{1}{\sqrt{1+Q^2}}}\delta^5 \, \text{cos}\,\alpha\frac{d\, \text{cos}\,\alpha }{d\,\text{cos}\,\psi}\frac{dA'}{D^2}I'_0 (\alpha'),
\end{equation}
where $A_s$ is the conformal factor evaluated on the stellar surface, $\delta$ is the Doppler factor (whose expression can be seen in \cite{SilvaeNunesPhysRevD.99.044034}), which takes into account the gravitational redshift, $\alpha$ is the local emission angle, $dA'$ is the area element on the surface, $D$ is the distance to the pulsar and $I_0$ is the specific intensity of radiation, which is naturally expressed in terms of $\alpha'$, the emission angle for an observer co-moving with the surface.

The values of $A_s$ and $Q$ for a given mass and compactness are needed to integrate the flux formula. With the mass, we can get the values of $b$ and $a$ using the scalar charge $Q$. The mass and compactness also specify the physical radius, which can be translated to $\rho_s$ via the conformal factor $A_s$. We can perform an EOS-independent analysis using the exterior spacetime solution \cite{SilvaeNunesPhysRevD.99.044034}. Namely, we can specify the mass and radius without integrating the interior relativistic structure equations and choosing suitable values for the scalar charge $Q$ and conformal factor $A_s$, thus characterizing the exterior spacetime. We adopt a straightforward and intuitive approach of treating the stellar compactness as an independent parameter, which maintains theoretical consistency within observational limits and the predictions of STT and GR. This method allows for a flexible yet robust exploration of the differences between the theories, clearly demonstrating how compactness influences the phenomena of interest without requiring an explicit TOV (or modified TOV) solution. Table \ref{tab1} lists the stellar models studied here.

To motivate possible compactness values to explore, we first present a specific solution following the model-dependent approach of integrating the interior equations for a given EOS to get the mass, radius, conformal factor at the surface, and the scalar charge via the asymptotic behavior of the external scalar field solution, Eq. (\ref{external scalar field}). We choose a realistic EOS and an STT model, like the ones presented in \ref{models}. These models depend generally on just one free parameter, the value of the coupling constant $\xi$. Once we choose the value of $\xi$ and a central pressure or density, we can integrate the equations outwards and obtain all the quantities needed to describe the exterior spacetime and compute the flux. As an example, Fig. \ref{compacteness} shows the equilibrium solutions for GR and STT ($\xi=-3$, $25$) for the realistic ENG \cite{engeos} and MPA1 \cite{mpa1eos} EOSs. GR and STT solutions have stellar configurations that go into the high compactness region to produce multiple images before reaching the maximum mass. One interesting case of specific coupling will be the one with $\xi>0$ since they \textit{scalarize} for stars with high mass and are still unconstrained by pulsar timing \cite{mandesandortizPhysRevD.93.124035,raissaTulioPhysRevD.99.124003}. Unfortunately, in this case, the smallness of the scalar charge makes the difference in the lightcurve relative to the GR case negligible \cite{raissaTulioPhysRevD.99.124003}. The differences between theories increase for high compactness, close to $M/R=0.284$, especially for the tangentially emitted photons $\alpha\approx \pi/2$ (meaning $\psi \approx \psi_c$), as can be seen in Fig. \ref{fig1}. 

The above solutions to the NS interior equilibrium equations with realistic EOSs demonstrate that GR and STT can result in high compactness configurations, allowing for multiple images in pulsars. To maintain generality, from now on, we adopt the previously mentioned model strategy of treating compactness as an independent parameter, having as references the critical threshold predicted by GR.

For the models considered here, following the choices of \cite{SilvaeNunesPhysRevD.99.044034} with $Q=0.5$ and a conformal factor varying $5\%$ relative to $1$, the critical angle $\psi_C$ becomes $\pi$ for a compactness in the range $[0.275,0.305]$. This slight difference creates a large qualitative difference in the lightcurve because a star with $\psi_C > \pi$ develops a region behind it that produces multiple images so the photon can bend from one direction or another. Meanwhile, for a $\psi_C < \pi$, there is an invisible zone behind the star where the photons cannot reach the observer.

To make the model appropriate for astrophysical applications, one must go beyond the infinitesimal approximation, integrating the flux over an extended region of the star's surface. In this case,
\begin{equation}\label{fluxformulacomplete}
    F = F_0 A_s^2(1 - \Bar{a}_s)^{\frac{1}{\sqrt{1+Q^2}}} \iint_S \delta^5 \text{cos}\alpha \,\text{sin}\alpha \frac{d\alpha}{d\psi}\,d\psi d\phi,
\end{equation}
where we choose spherical coordinates over the spot area $S$. Since we are dealing primarily with compact configurations, where the view bending angle can become larger than $\pi$, it is better to write the derivative as $\frac{d\alpha}{d\psi}$ to avoid the singularity when $\text{cos} \psi=0$. Also, $F_0$ is a phase-independent overall constant
\begin{equation}
    F_0 \equiv \frac{I_0}{R D^2}. 
\end{equation}

Owing to the exploratory theoretical nature of the present work, we consider only one hot spot on the stellar surface, with a circular shape of semi-aperture angle $\Delta\psi$, to isolate the scalar-field effects. Likely, the light-curve fitting of specific sources could require additional ingredients, such as complicated magnetic field structures over a simple-centered dipole or multiple spots \cite{Bilous2019,magneticfield10.1093mnrasstv598,Rafael2022arXiv221006648D}. The infinitesimal approximation works well for small spots ($\Delta\psi<5\degree$). For larger spots, one must integrate over the spot area. In this case, the difference between the theories increases because of the cumulative light bending, time delay, and gravitational redshift.

Let us briefly revisit the massive STT mentioned earlier. A high scalar field mass always suppresses the scalarization of the star, leading to a lower scalar field value at its surface \cite{massivescalarPhysRevD.93.064005}. For example, if one assumes that the flux equation \eqref{fluxformulacomplete} approximately holds for the massive case, Fig. 2 of Ref. \cite{massivescalarPhysRevD.93.064005} suggests that for $m_{\varphi}>1.6\times 10^{-12} eV$, the flux differences between STT and GR become negligible. Therefore, the results presented in the next section for the massless case can be regarded as an upper limit on flux changes (see also \cite{highlymassivePhysRevD.96.084026} for a discussion of a massive scalar field in heavy NS with the ``asymmetron" model). We leave precise details about the flux change in the case of highly compact NS with a massive scalar field for future work.

\begin{figure}
\centering
\includegraphics[width=0.8\columnwidth]{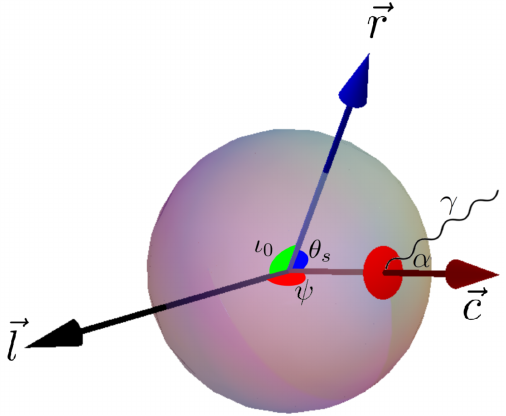}
    \caption{Schematic representation of angles and vectors for the NS. \(\vec{l}\) represents the line of sight, \(\vec{r}\) denotes the axis of rotation, and \(\vec{c}\) is normal to the center of the polar cap. The angle \(\psi\) is defined between \(\vec{l}\) and \(\vec{c}\), and \(\alpha\) represents the angle made by a photon leaving the star relative to \(\vec{c}\). Additionally, \(\iota_0\) is the angle between \(\vec{l}\) and \(\vec{r}\), and \(\theta_s\) is the angle between \(\vec{c}\) and \(\vec{r}\).}

\label{fig:star_angles_vectors}
\end{figure}

To characterize the flux, one must know the specific geometry of the source, which can be described by the angles $(\iota_0,\theta_s)$, as illustrated in Fig. \ref{fig:star_angles_vectors}. Here, $\iota_0$ is the angle between the rotational axis of the NS and line of sight (LoS), and $\theta_s$ is the colatitude of the spot's center relative to the rotational axis. The position of the spot's center will vary in time as the star rotates
\begin{equation}
    \text{cos}\,\psi_0 = \text{sin}\,\iota_0\,\text{sin}\,\theta_s\,\text{cos}\,\omega t + \text{cos}\,\iota_0\,\text{cos}\,\theta_s,
\end{equation}
where $\omega$ is the angular velocity of the star, and we choose $t=0$ as the moment of closest approach between the spot and the observer. For the integration procedure, we follow \cite{Turolla_2013,deLima2020,Rafael2022arXiv221006648D}. The main difference here is the inclusion of the flux of the secondary image of the spot. The center of the secondary image is in the position $\psi_{sec}=2\pi-\psi_0$, and we start to consider it when $\text{cos}\,(\psi_0 + \Delta \psi) \leq \text{cos}\,(\psi_c) $.

Here, we keep things simple enough to isolate the effects of STT on the lightcurves of compact NS. We do not include an atmosphere model, magnetic fields, and rotation effects on the exterior spacetime and star's structure. These ingredients must be included in STT \cite{PaniBertiPhysRevD.90.024025} to be consistent.

\begin{table*}[h!] \label{table}
    \centering
    \begin{tabular}{ |p{1cm}|p{0.8cm}|p{1.1cm}|p{1cm}|p{1cm}|p{0.8cm}|p{0.8cm}|p{0.8cm}|p{0.6cm}|}
 \hline
 \multicolumn{9}{|c|}{Stellar Models} \\
 \hline
 Name & $M/R$ & $M/M_{\odot}$ & $R/\rm km$ & $\rho_s/\rm km$ & $A_s$ & $\Bar{a}_s$  & $\Bar{b}_s$ & Q \\
 \hline
   GR &  0.284   &   2.1  & 10.918& 10.918    & 1 & 0.568   & 0.568 & 0  \\
   STT1 & 0.284  &   2.1 & 10.918 & 12.026 & 0.95     & 0.576  & 0.515 & 0.5 \\
  STT2 &  0.284 &   2.1  & 10.918 & 10.962 & 1.05 & 0.632    & 0.565 & 0.5  \\
 \hline
\end{tabular}
    \caption{Stellar models considered in the light-curve analysis. We choose configurations that are doppelgängers of each other, with the same mass M and Jordan frame radius R, giving the same compactness value of $M/R=0.284$. The STT models are chosen by fixing the conformal factor at the surface $A_s$ and scalar charge $Q$. The values of $\rho_s$ (Einstein Frame radius in Just coordinates), $\Bar{a}_s = a/\rho_s$ and $\Bar{b}_s = b/\rho_s$ are then obtained by Eq. \ref{radialcoordinates}, the b definition and Eq. \ref{constraintcharge}, respectively.}
    \label{tab1}
\end{table*}

\section{Results} \label{results}

\begin{figure}
\centering
\includegraphics[width=\columnwidth]{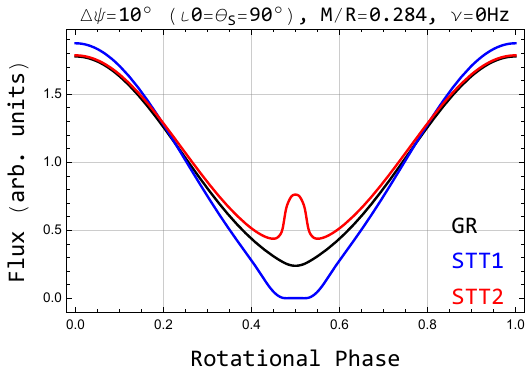}
\caption{Bolometric flux of the models relative to GR. The difference is more evident when the spot passes the area opposite the observer's line of sight, around half the rotational period. In one case there is an invisible zone and a multiple image zone in the other, producing a brightening in the flux. The huge difference between the fluxes is evident, both qualitatively and quantitatively. }
\label{fig2}
\end{figure} 
 
In Fig. \ref{fig2}, we show the bolometric flux of a spot with $\Delta \psi = 10 \degree$ for the compact stellar models of Table \ref{tab1}. In this case, we do not consider rotational effects (Doppler and time delay) and take the rotational frequency $\nu=0$. For the GR star, the critical angle is $\psi_c = \pi$, and the whole surface of the star is visible, making a non-vanishing flux, even when the spot is behind the star. For the model STT1, according to Fig. \ref{fig1}, the critical angle is smaller than $\pi$ ($\psi_c=0.90\pi$), making an invisible zone of roughly $\sim 20 \degree$ behind the star, which eclipses the entire spot during a short period. In the case of the model STT2, the critical angle is a little greater than $\pi$ ($\psi_c=1.05\pi$), meaning that a zone of roughly $\sim 9 \degree$ starts to produce a second image when the spot reaches it, making that brightening observed in the lightcurve close to half the rational phase of the star.

One interesting fact about the light bending of compact stars is that visually, the star appears bigger than it is. For a $2.1M_\odot$ star, with compactness $M/R=0.284$, the physical radius is  $10.918~\rm km$. Still, using the impact parameter formula (\ref{impact parameter formula}) for the last photon that we can see from the star surface $(\alpha=\pi/2)$, we find a value of $16.611~\rm km$ for GR, almost $50\%$ bigger. A different gravitational field will of course also influence the visual appearance of the star, for the STT1 model the apparent radius is $\sim260~m$ bigger, and for STT2 $\sim340~m$ smaller, causing although a small perceptual difference of $1.6\%$ and $2\%$ respectively, with respect to GR.

These distinct features between the light curves can be appreciated for increasing compactness above $M/R=0.275$ (see Appendix \ref{apB}) and a geometrical configuration where the spot crosses the invisible or multiple image region behind the star. To be more specific, the effect is evident when $\iota_0+\theta_s - \pi \leq \Delta\psi$, meaning that the ideal configuration to test is with small hot spots seen edge-on that are near the stellar equator. Although it may sound particular, the analysis of Miller et al. \cite{2021ApJ...918L..28M} of NICER data from the massive ($2.08 M_{\odot}$) pulsar PSR J0740+6620 is consistent with this geometrical configuration, making this source suitable for gravitational theory tests.

\textbf{}

Although the lightcurves are qualitatively different for the same compactness, there is a degeneracy between the compactness and the scalar charge: a slightly more compact star in GR will start to produce the multiple image region, and a less compact one will produce an invisible zone. This is a reflex of the well-known degeneracy between the equation of state physics and the gravitational theory \citep[see, e.g.,][]{degeneracy10.1063/1.5117806}. To get a clear signature of the scalar charge on the lightcurve, one, in principle, needs an independent measurement of the mass and radius, which is difficult for millisecond pulsars, for example. The mass can be well measured by the radio timing of the pulses, while the radius measurement is way more elusive \cite{ASCENZI2024102935}. Still, multimessenger observations could help to break this degeneracy.

\begin{figure}[h]
\centering
\includegraphics[width=\columnwidth]{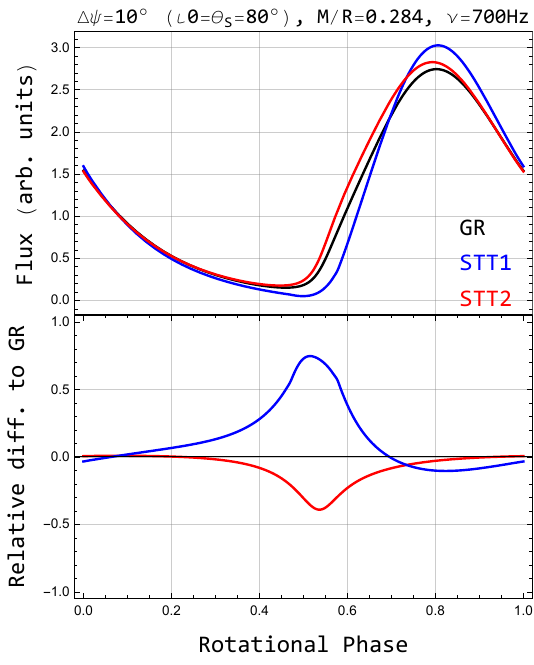}
\caption{Upper: Bolometric flux for a geometric configuration $\iota_0=\theta_s=80\degree$, where the spot's flux does not suffer the influence of the special region behind the star, so the differences are ascribed to the time-delay and gravitational redshift. Lower: STT cases relative difference relative to GR.}
\label{fig3}
\end{figure}

Even for geometric configurations where the spot does not cross the region opposite to the line of sight, the lightcurves of the models can be significantly different when we include rotational effects. Here, the spacetime is still spherically symmetric, and we keep the spherical shape of the star. Still, we include special relativity effects that depend crucially on the gravitational redshift and the time delay, which are small but enhanced by the integration over the extended spot. Motivated by observations, we choose, in the case of rotation, the frequency $\nu=700$ Hz. In Fig \ref{fig3}, we show the bolometric lightcurve for a $\Delta \psi = 10 \degree$ spot but for a configuration \textit{almost equatorial}, with $\iota_0=\theta_s=80\degree$. In this case, we do not have the effect illustrated in Fig. \ref{fig2} because the spot does not cross exactly the region behind the star. The difference is mainly due to the surface gravity that affects the special relativistic effects and the time delay integrated over the extended spot.

We emphasize that for a fast-rotating NS, one should include at least the dominant effect of the deformed stellar surface in the lightcurve analysis \cite{Cadeau_2007}, but for geometrical configurations where $\iota_0 \approx \theta_s \approx 90\degree$, as discussed here, the approximation using a spherical surface works well and the effects of a rotating spacetime is subdominant, as demonstrated by Cadeau et al. \cite{Cadeau_2007} for the GR case.

\section{Conclusion}\label{sec13}

While GR has passed all experimental tests, it remains imperative to scrutinize it for potential deviations, striving for greater precision. Among the notable alternative theories is Scalar-Tensor gravity, which posits an additional scalar degree of freedom alongside the metric tensor to describe gravitation and predict different phenomena, like \textit{spontaneous scalarization} that affect both neutron stars structure and gravitational field. 

In this work, we analyze a promising way to test STTs in the strong-field regime of high-mass pulsars: the pulse profile is very sensitive to the scalar charge when the compactness is close to $GM/(c^2R) \sim 0.284$. It makes compact systems with localized hot spot emission close to the equator, ideal for testing scalar-tensor theories and suggestively,  alternative gravitational theories in general. Such an analysis needs the inclusion of all ingredients that enter the pulse profile analysis: gravitational redshift, light bending, time delay, and different stellar structures. The advantage of the STT used here is that it allows the analysis with an analytic, closed-form exterior spacetime solution for the light bending and time delay. The STT effects, especially for the light bending, are more pronounced for the light rays emitted tangentially relative to the stellar surface, as shown by the distinct features in Figs. \ref{fig2} and \ref{fig3}. The latter shows flux differences of up to $80\%$ in the spot's passage around the region opposite the line of sight (i.e., behind the star relative to the observer). This suggests that differences between GR and STT lightcurves can be significant in cases of high compactness. Therefore, that could be a promising observational approach to constrain deviations from GR.  Due to the nonlinear dependence of the flux on compactness, slightly smaller values of the latter can result in much smaller relative changes of the former in STT and GR. 

Figure \ref{fig1} shows the nonlinearity of the differences between STTs and GR in the deflection angle. For a fixed scalar charge, configurations with a scale factor with the same excess or defect relative to the GR case lead to asymmetric stellar compactness values (relative to the GR case) at which the whole NS becomes visible. This is expected because STT changes are more relevant for more relativistic systems. The largest deviation of compactness is around $10\%$ (when the star is not entirely visible $(\psi_c<\pi)$, the relative changes are smaller). It is meaningful to compare the above numbers with NICER compactness constraints, which use GR, to gain insight into the feasibility of probing scalarization in compact stars. We use as reference pulsar PSR J0740+6620, constrained to having a radius \citep{2021ApJ...918L..27R,2021ApJ...918L..28M} $R=12.39^{+2.22}_{-1.50}$ km (90$\%$ credible interval). For the accurately-estimated mass of $2.08 M_{\odot}$, its compactness is $0.209$--$0.281$, with a median of $0.248$. Thus, the relative dispersion of compactness values is around $12\%$--$15\%$. Therefore, current NICER observations cannot yet differentiate GR from STTs. However, with the increasing accuracy of multimessenger constraints \citep{2021ApJ...918L..28M} or several gravitational-wave observations \citep{2022PhRvD.105h4021C}, it will be possible to differentiate theories using ray-tracing observables, especially for highly compact stars. The largest compactness dispersion produced by STT can also be used to estimate the radius uncertainty associated with scalarization. From the definition, ${\cal C}\equiv M/R$, it follows that $|\Delta R|/R=\Delta {\cal C}/{\cal C}$ for a well-constrained mass. Thus, $|\Delta R|/R\lesssim 10\%$ for the above parameters. This also suggests that $Q=0.5$ is the maximum value of the charge parameter allowed by current radius constraints. Naturally, tighter constraints on $Q$ and $A$ or smaller values for $\psi_c$ will reduce that uncertainty. Still, it is large enough to suggest that STT could generally impact radius inferences from lightcurves.

Large compactness values are generally associated with very dense systems where phase transitions can occur \citep{Annala:2019puf,2022PhRvD.105l3015P,2023ApJ...955..100G}. This means that probing scalarization could also be particularly relevant in hybrid stars \citep{2018ApJ...860...12P}. If the surface tension of dense matter is large enough, the quark phase and the hadronic phase could be in direct touch (first-order phase transition). In this case, phase conversions could happen upon perturbations, and the matter would not be catalyzed anymore, meaning that the usual stability rules for NSs would be violated \citep{2018ApJ...860...12P}. Such a violation would allow for an extended branch of (meta)stable NSs with large compactnesses (see, e.g., \citep{2020PhRvD.101j3003D,2020PhRvD.101l3029T,2021Univ....7..493L,2022EPJC...82..288G,2023PhRvD.107j3042R}). The terminal mass (where radial perturbations have null eigenfrequencies) within this branch is not known. Still, it could go down to values about ordinary stars \citep{2021ApJ...910..145P}. All the above means that scalarization could be relevant for NSs of canonical mass, in addition to massive ones.

Gravitational theory tests using pulse profile modeling are still not sufficiently competitive compared to binary pulsar experiments for real astrophysical verification or for placing meaningful constraints. However, this subject remains a fresh and fertile area for research, as emphasized in \cite{silva_neutron_2024}. In this work, we have presented promising configurations that could shed new light on the field. However, our model is still too simplified for application to real astrophysical sources. We must incorporate the possibility of multiple spots (e.g., \cite{deLima2020,Rafael2022arXiv221006648D}) with temperature distributions and account for atmospheric effects within the context of STT, which can attenuate tangentially emitted photons. Additionally, it is crucial to consider stellar oblateness. For instance, as an initial approximation, one could solve the structure equations up to second order in angular velocity \cite{PaniBertiPhysRevD.90.024025} to obtain the star's quadrupole moment and shape, resulting in an oblate Just+Doppler model for the NS spacetime. With this more comprehensive theoretical model in hand, a statistical analysis can be performed, for example, using the NICER data for the high-mass pulsar PSR J0740+6620. This would enable us to obtain posterior values for the mass and radius and the parameters of the STT. We leave this analysis for future work.

\backmatter

\bmhead{Acknowledgements}
We thank the anonymous referee for their valuable suggestions and comments, which have helped improve this work.
T. O. is grateful for the financial support of CAPES and the hospitality of the University of Ferrara. J.G.C. is grateful for the support of Funda\c c\~ao de Amparo \`{a} Pesquisa e Inova\c c\~ao do Esp\'irito Santo (FAPES) (grant Nos. 1020/2022, 1081/2022, 976/2022, 332/2023), CNPq (grant No. 311758/2021-5), and Funda\c c\~ao de Amparo \`{a} Pesquisa do Estado de S\~ao Paulo (FAPESP) (grant N. 2021/01089-1). R.C.R.L. acknowledges the support of Funda\c c\~ ao de Amparo \` a Pesquisa e Inova\c c\~ ao do Estado de Santa Catarina (FAPESC) under grant No. 2021TR912.
J.P.P. is thankful for the financial support of FAPES under grant No. 04/2022. This study was partly funded by the Coordena\c c\~ ao de Aperfei\c coamento de Pessoal de N\' ivel Superior – Brasil (CAPES)--Finance Code 001.

\begin{appendices}

\section{The Beloborodov Approximation}\label{belo aprox}

\setcounter{figure}{5}

Although this work focuses on the lightcurve of compact stars, a comment can be made about the well-established Beloborodov analytical approximation \cite{Beloborodov2002} in the context of STT. This approximation for the light bending integral, Eq. \ref{bendingintegral}, is valid for compactness bellow $M/R\leq0.25$, where we get an exact analytical formula without solving the integral numerically, saving computational time.

\begin{figure}[h!]
\centering
\includegraphics[width=\columnwidth]{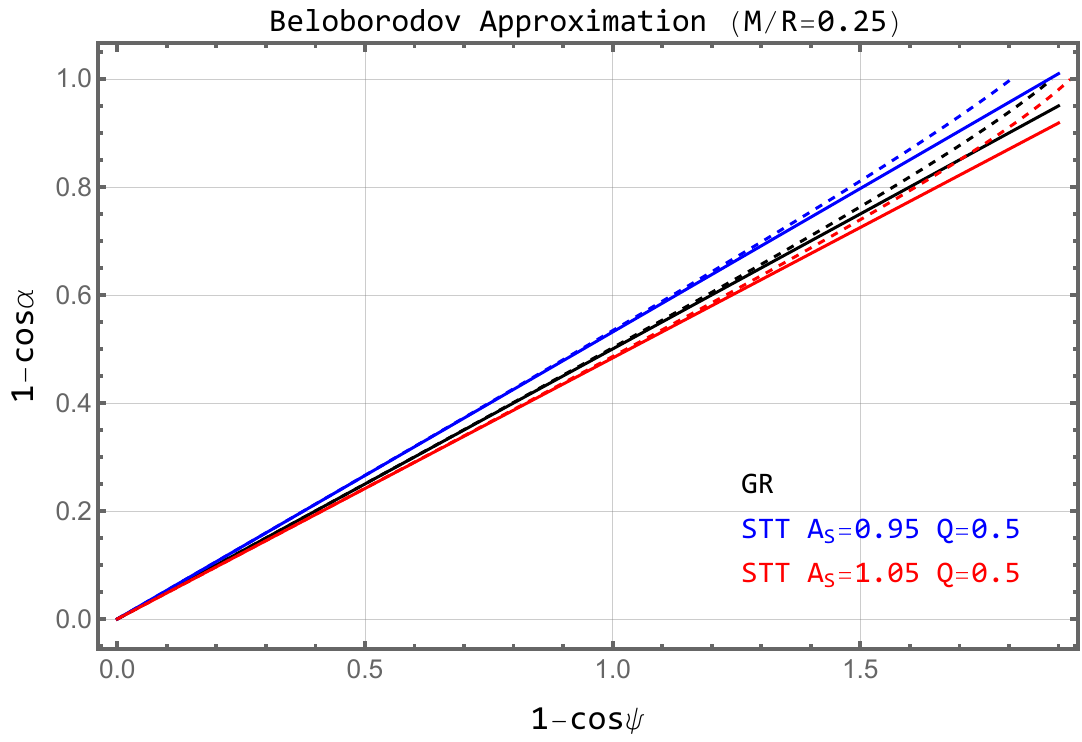}
\caption{Beloborodov approximation for compactness of $M/R=0.25$. The dashed lines are the exact numerical result, and the solid lines are for the function $(1-\text{cos}\alpha) = (1-2x)(1-\text{cos}\psi)$, with $x=0.25$ for GR and $0.234$, $0.258$ for the STT models. The fact that the lines run practically parallel shows compactness's main role in this geometric approximation. The error is a maximum of $7\%$ for the emission angle $\alpha = \pi/2$.}
\label{belobo}
\end{figure} 

The Beloborodov formula reads
\begin{equation}
    1-\text{cos}\alpha=(1-2{\cal C}) (1-\text{cos}\psi)
\end{equation}
where ${\cal C}$ is the stellar compactness. In Fig. \ref{belobo}, we show the Beloborodov approximation for GR and the equivalent for the STT, choosing the value of the slope so that the error relative to the numerical results is similar. We fixed the scalar charge to be $Q=0.5$ and the scale factor at the star's surface to vary $5\%$ relative to $1$. For the blue curve ($A_s=0.95$), we put ${\cal C}=0.234$, and for the red one ($A_s=1.05$) ${\cal C}=0.258$. We note that these values are very similar to the  ``effective compactness" in the STT solution $b_s/2=GM/\rho_s c^2$, which are $b_s/2=0.228$ and $b_s/2=0.251$. We can interpret this as the compactness's key role in light bending, which dominates over the direct effect of the scalar charge \cite{Sotani1PhysRevD.96.104010}. The lines of GR and STT run almost parallel, and the error is a maximum of $7\%$ for the emission angle $\alpha=\pi/2$.

\section{Increasing compactness and the appearance of multiple images of the spot}
\label{apB}

\begin{figure}[h!]
\centering
\includegraphics[width=\columnwidth]{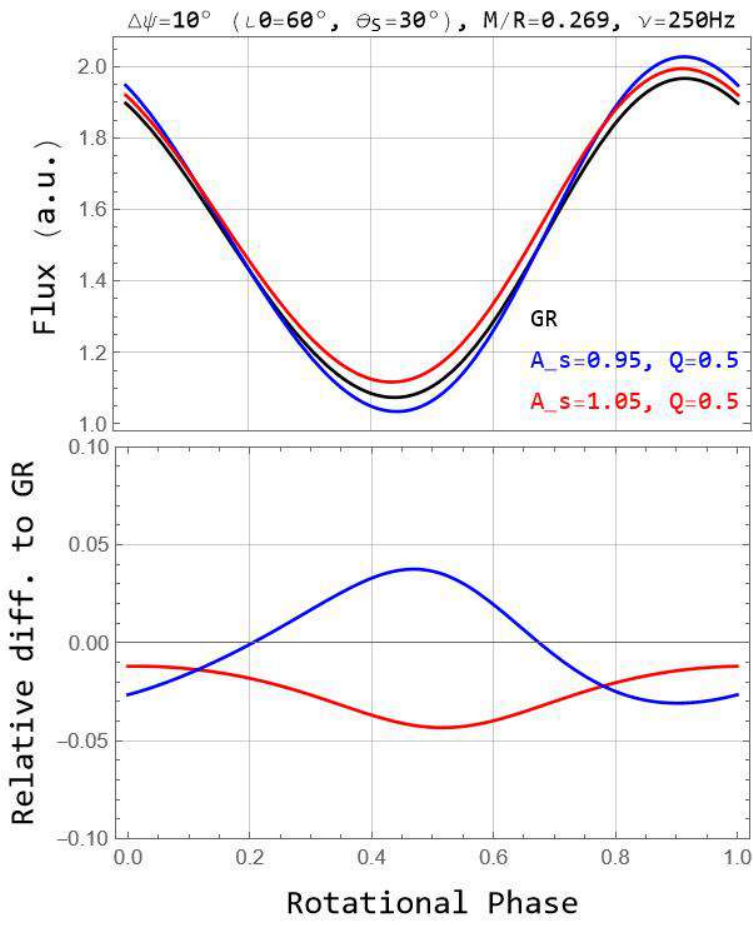}
\caption{The difference between the theories is small for compactness below critical and a geometrical configuration where the spot does not cross the region opposite to the LoS.}
\label{low compactness}
\end{figure}

One of the key aspects of the results discussed here is the appearance of multiple images of the NS surface when the compactness increases over a critical value, which depends on the gravitational theory and crucially affects the lightcurve. The difference between the theories is small for compactness below the critical one, as seen from Fig. \ref{low compactness}. The geometric configuration is similar to those of Fig. 5 of Hu \textit{et al.} \cite{HuGaoXuShaoPhysRevD.104.104014}, and we also find within our model that the maximum relative change of the flux in STT and GR is smaller than $5\%$, in agreement with their findings.

\begin{figure}[h!]
\centering
\includegraphics[width=\columnwidth]{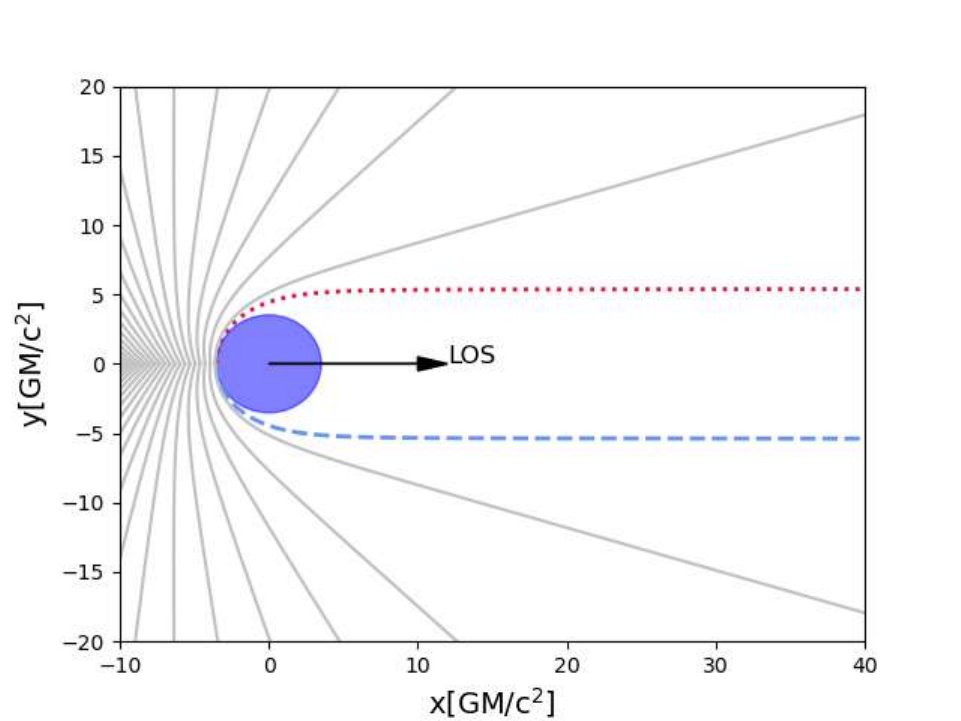}
\caption{Several geodesics starting from different heights relative to the NS equator are shown from an overhead view, with the line of sight (LOS) along the x-axis. The lensing effect of extreme bending is evident in two geodesics—red dotted and blue dashed—that originate from the same point behind the star's surface ($\psi=\pi$) and still reach the observer ($\psi=0$) via two different paths. That is related to multiple images when the compactness is large enough. The NS compactness is ${\cal C} = 0.284$, and the spacetime is Schwarzschild.}
\label{photon path}
\end{figure}

But as compactness increases, the bending becomes strong enough so that the critical angle $\psi_c$ can be greater than $\pi$, meaning that photons that leave a region behind the star, relative to the LoS, can take two paths to reach the observer, a phenomenon similar to gravitational lensing that can be seen in Fig. \ref{photon path}. We consider the Schwarzschild spacetime and a star with the critical compactness ${\cal C} = 0.284$.

\begin{figure}[h!]
\centering
\includegraphics[width=\columnwidth]{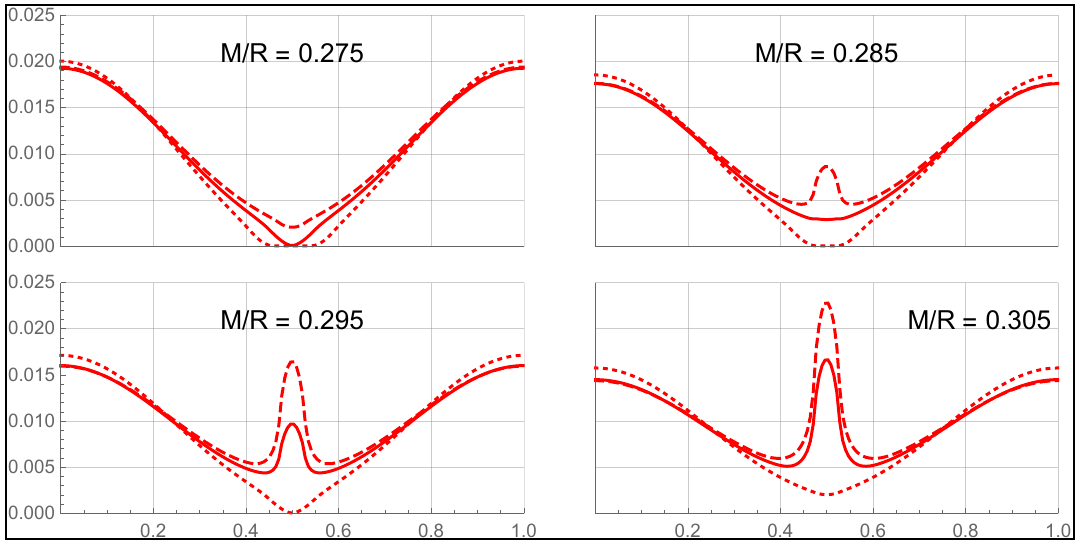}
\caption{{Bolometric flux fixing the geometrical configuration to be $\iota_0=\theta_s=90\degree$ with a spot size $\Delta\psi=10\degree$. The GR light curve is the thick one, while for the dotted one, we set $A_s=0.95, Q=0.5$ and for the dashed $A_s=1.05, Q=0.5$.}}
\label{plot lightcurve compact}
\end{figure}

One can appreciate in Fig. \ref{plot lightcurve compact} the increasing sensitivity of the light curve as compactness increases in the interval 
$[0.275,0.305]$, where the GR solution (thick line) can be very different from the STT models considered (dotted and dashed). The brightening observed around half the rotational period, associated with the lensed secondary flux, can be compared with \cite{Sotani3PhysRevD.98.044017,Sotani4PhysRevD.101.063013}. The peak of the brightening depends on the spot size, with smaller spots producing more pronounced peaks \cite{Sotani4PhysRevD.101.063013}.

\end{appendices}

\bibliography{sn-bibliography}

\end{document}